\documentclass[twoside]{ae100prg}
\usepackage{amsmath}
\usepackage{graphicx}
\usepackage[breaklinks]{hyperref}
\usepackage{booktabs}

\begin{document}

\title{Modified gravity theories and dark matter models tested by galactic rotation curves}
\author{Marek Dwornik$^{1}$, Zolt\'{a}n Keresztes$^{1}$, L\'{a}szl\'{o} \'{A}rp\'{a}d Gergely$^{1}$}
\email{marek@titan.physx.u-szeged.hu}

\begin{abstract}
Bose-Einstein condensate dark matter model and Randall-Sundrum type 2 brane-world
theory are tested with galactic rotation curves. Analytical expressions are derived
for the rotational velocities of test particles around the galactic center
in both cases. The velocity profiles are fitted to the observed rotation
curve data of high surface brightness and
low surface brightness galaxies. The brane-world model fits better the rotation curves with
asymptotically flat behaviour.

\end{abstract}

\address{$^1$ University of Szeged, Departments of Theoretical and Experimental Physics, D\'{o}m t\'{e}r 9. 6720 Szeged, Hungary}

\section{Introduction}

Luminous matter alone can not explain the observed behaviour of the galactic
rotation curves and an invisible, dark matter component is needed \cite{persic96}. 
One possibility to explain dark matter is to introduce collisionless dark
scalar particles in form of a Bose-Einstein condensate (BEC) \cite{sin94}, \cite{harko07}. 
However, up to now the evidence for dark matter has been only found by its gravitational
interaction. It cannot be excluded that general relativity breaks down at
scales of galaxies and beyond. Therefore several modified gravity models
have been proposed to replace dark matter \cite{milgrom83}, \cite{sanders84}, \cite{roberts04}. 
The Weyl fluid appearing in Randall-Sundrum type 2 (RS2) brane-world models can behave as an 
effective source for gravity and it is able to replace dark matter in galactic dynamics \cite{mak04}, \cite{gergely11}.

We investigate here galactic rotation curves in RS2 brane-world and the BEC model for high surface brightness (HSB) and low surface brightness (LSB) galaxies.

\section{The baryonic matter}

Distribution of baryonic matter in HSB galaxies is
described as the sum of a thin stellar disk and a spherically symmetric
bulge component.

We assume that the mass distribution of bulge component with radius $r_{bulge}$ follows the
de-projected luminosity distribution with a factor called mass-to-light
ratio $\sigma $. The surface brightness profile of the spheroidal bulge
is described by a generalized S\'{e}rsic function 
\cite{sersic68}: $I_{bulge}(r)=I_{0,bulge}\exp \left[ -\left( r/r_{0}\right)
^{1/n}\right] $, where $I_{0,bulge}$ and $r_{0}$ are the central surface
brightness and the characteristic radius of the bulge, respectively, and $n$
is the shape parameter of the magnitude-radius curve. The contribution to
the rotational velocity is
\begin{equation}
v_{bulge}^{2}(r)=\frac{GM_{bulge}(r)}{r},  \label{vbulge}
\end{equation} 
with gravitational constant $G$ and mass of the bulge $M_{bulge}(r)=\sigma I_{bulge}(r)$ within the radius $r$ 
projected on the sky.

In a spiral galaxy the radial surface brightness profile of the disk decreases 
exponentially with the radius \cite{freeman70}: $I_{disk}(r)=I_{0,disk}\exp \left( -r/h\right) 
,$ where $I_{0,disk}$ is the disk central surface brightness and $h$ is a
characteristic disk length scale. The contribution of the disk to the
circular velocity is \cite{freeman70}: 
\begin{equation}
v_{disk}^{2}(x)=\frac{GM_{disk}}{2h}x^{2}(I_{0}K_{0}-I_{1}K_{1}),
\label{vdisk}
\end{equation}
where $x=r/h$ and $M_{disk}$ is the total mass of the disk. The
functions $I_{m}$ and $K_{m}$ are the modified Bessel functions of the first
and second kind with order $m$, respectively. The Bessel functions are evaluated at $x/2$.

The best fitting values of  $I_{0,bulge},n,r_{0},r_{bulge},I_{0,disk}$
and $h$ are derived from the available photometric data. In case of LSB galaxies the
baryonic model only consists of a thin stellar disk component which is the
same as at the HSB galaxies. 

\section{Models for the dark matter component}

The mass density distribution of the static gravitationally bounded Bose-Einstein condensate
is described by the Lane-Emden equation. An analytical solution for dark matter BEC was obtained 
in Ref. \cite{harko07}: $\rho _{BEC}(r) =\rho_{BEC}^{(c)} \left( \sin (kr)/kr\right)$, 
where $k=\sqrt{Gm^{3}/\hbar ^{2}a}$ and $\rho _{BEC}^{(c)}$ is the central mass density of the condensate. 
The mass profile of the galactic halo is $m_{BEC}\left( r\right) =4\pi \int \rho _{BEC}(r)r^{2}dr$ 
giving the following contribution to the rotational velocity \cite{harko07}:
\begin{equation}
v_{BEC}^{2}(r)=\frac{4\pi G\rho _{BEC}^{(c)}}{k^{2}}\left( \frac{\sin kr}{kr}-\cos kr\right) .  \label{vel}
\end{equation}

In RS2 brane-world theory the 4-dimensional effective Einstein equation has
extra source terms, which arise from the embedding of the 3-brane in the
bulk \cite{shiro00}. We assume that the brane embedding is $Z_{2}$-symmetric and
there is no matter in the 5-dimensional spacetime but there is a cosmological
constant. Nevertheless the effect of the brane cosmological constant arising from the
brane tension and the 5-dimensional cosmological constant is neglected at
the scales of galaxies. Then at low energies there is only one extra source
term in the effective Einstein equation arising from the 5-dimensional
Weyl curvature, which acts as a fluid (the Weyl fluid). The contribution to the rotational velocity in
a Post-Newtonian approximation is derived in \cite{gergely11}:
\begin{equation}
v_{Weyl}^{2}(r)\approx {\frac{{G}\left( M_{0}^{tot}\right) }{{r}}}
+c^{2}\beta +c^{2}C\left( \frac{r_{b}}{r}\right) ^{1-\alpha }~,~r>r^{\ast },
\label{v2tg}
\end{equation}
with constants $\alpha $, $\beta $, $C$ and $M_{0}^{tot}$ charaterizing the
Weyl fluid and velocity of light $c$. A scaling constant $r_{b}$ was introduced such that C becomes 
dimensionless and $r_{b}=r_{bulge}$ was chosen. The rotational velocity (\ref{v2tg}) is
valid for any $r$ $>r^{\ast }$, where $r^{\ast }$ represents the radius
beyond which the baryonic matter does not extend. We assume that $i)$ the
contribution of the Weyl fluid can be neglected within the bulge radius, 
and $ii)$ the observed rotation curves within $r_{bulge}$ can be
explained with baryonic matter alone (it is given by the sum of Eqs. (\ref{vbulge}) 
and (\ref{vdisk}) for HSB and by (\ref{vdisk})\ for LSB galaxies). Outside $r_{bulge}$ we assume the effects of the disk 
can be handled as perturbation, therefore the rotational curves is generated by the sum of Eqs. (\ref{vdisk}) and (\ref{v2tg}).

\begin{figure}[h!]
\begin{center}
\begin{tabular}{cc}
\includegraphics [height=5.3cm, angle=270] {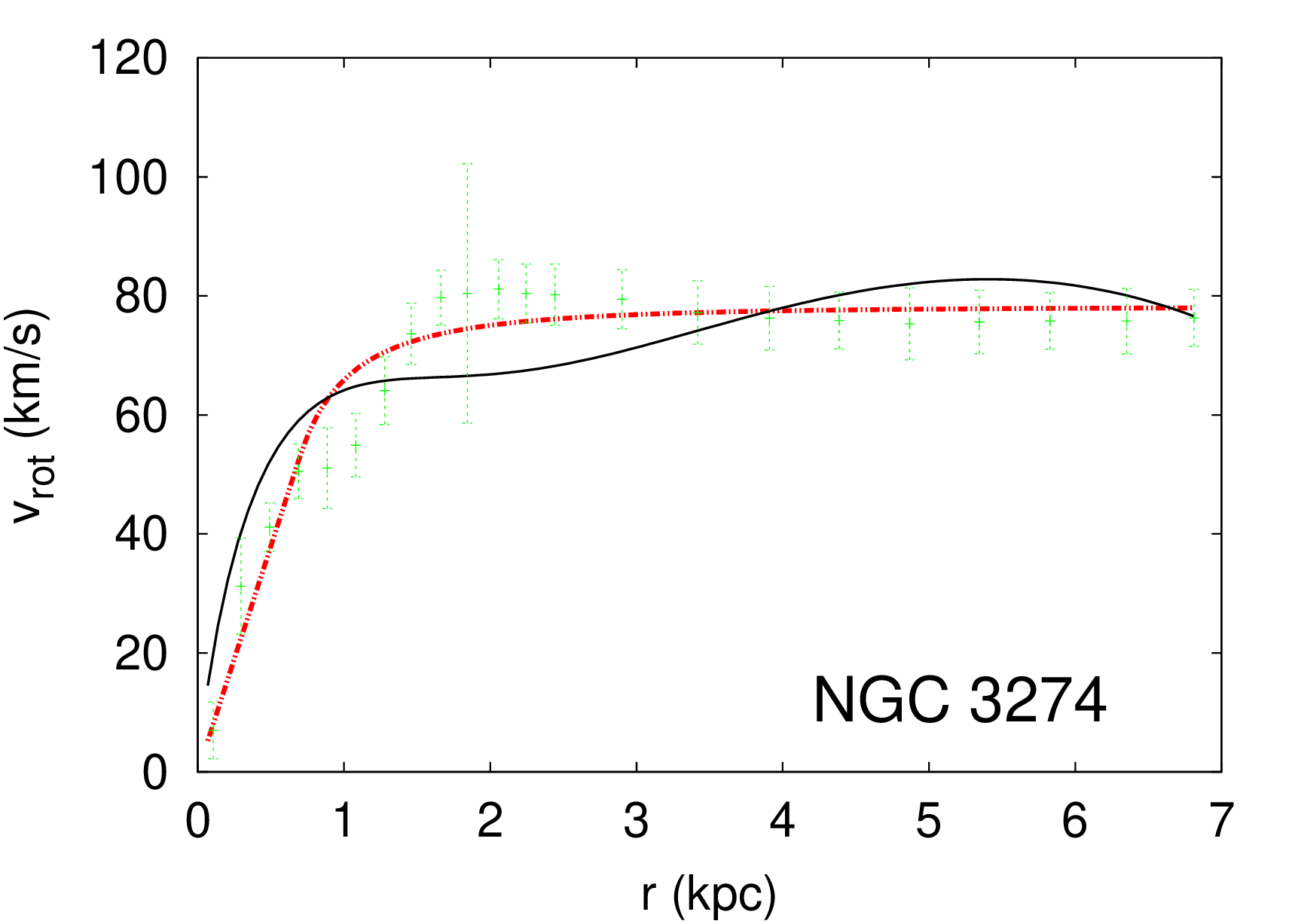} &
\includegraphics [height=5.3cm, angle=270] {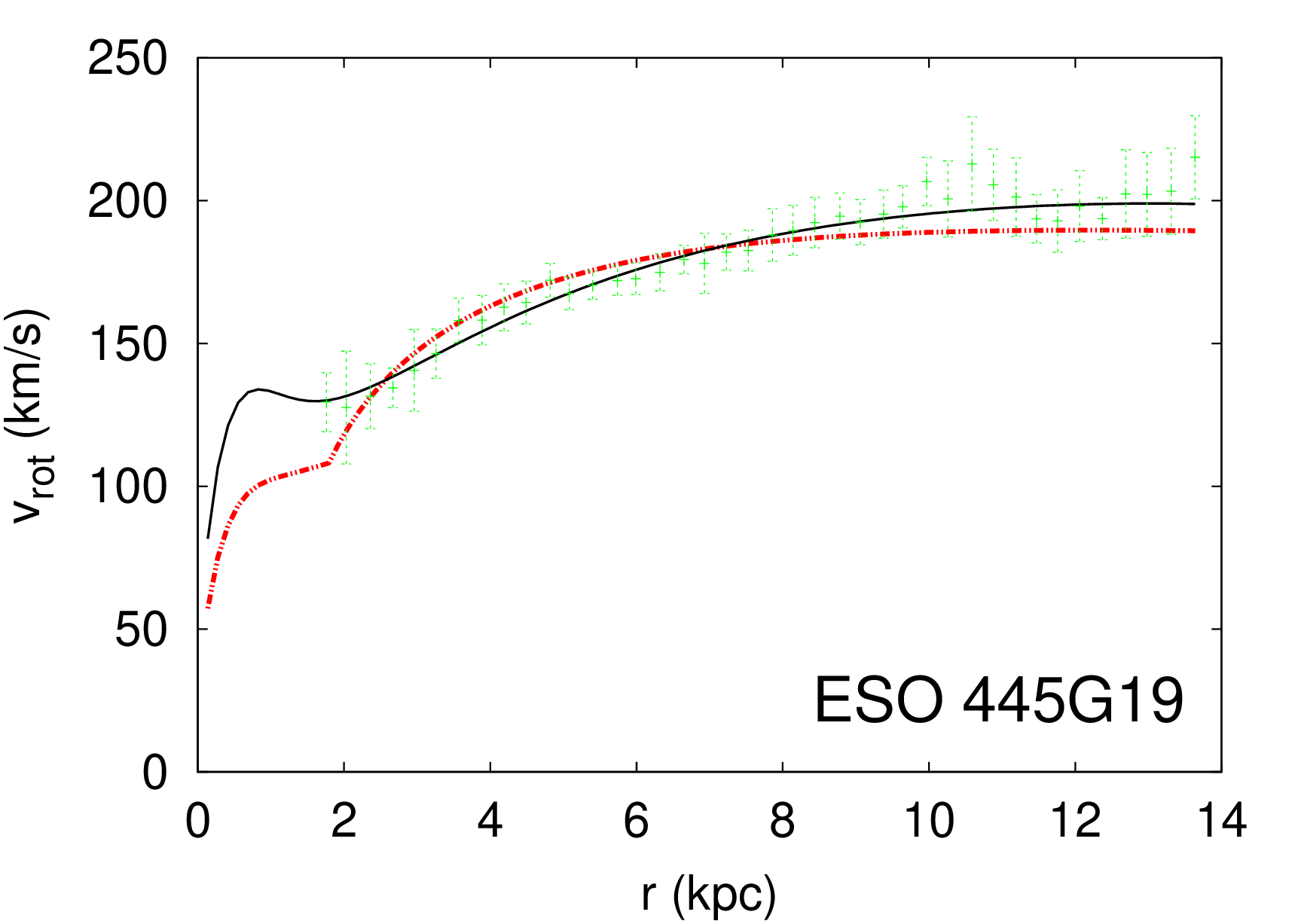} \\
\end{tabular}
\end{center}
\caption{Best fit curves for the observed velocity profiles of the HSB
galaxy ESO445G19 and the LSB galaxy NGC3274. The solid black and dashed red
curves show the BEC and Weyl models, respectively. In case of HSB galaxy, the shape of the curve 
near the center where baryonic matter dominates is
determined by the photometric data.} 
\label{fig01}
\end{figure}

The best fit rotational curves of the Weyl and BEC models for the observed
velocity profiles of the HSB galaxy ESO445G19 and the LSB galaxy NGC3274 are
shown on Fig. \ref{fig01}.

\section{Concluding Remarks}
We investigated whether RS2 brane-world and Bose-Einstein condensate
dark matter models can explaine the galactic rotational curves. Analytical
expressions for the rotational velocity of a test particle around the
galactic center in both model scenarios were derived. The rotation curves can
be well-explained by both models and we represented this for both a HSB and a LSB
galaxies on Fig. \ref{fig01}. The Weyl model was confronted with a larger
galaxy sample, finding good agrement with the observations in \cite{gergely11}.

\section*{References}

\bibliographystyle{plain}
\bibliography{mdwornik_prague}

\end{document}